\documentclass[aps,prb,twocolumn,groupedaddress]{revtex4-1}
\usepackage{amsmath}
\usepackage{amssymb}
\usepackage{amsbsy}
\usepackage{amsfonts}
\usepackage{amsopn}
\usepackage{amstext}
\usepackage{graphicx}
\usepackage{amssymb}
\usepackage{amsfonts}
\usepackage{amsmath}
\usepackage{graphicx}
\usepackage[english]{babel}
\usepackage{color}
\usepackage{slashed}
\usepackage{esint}
\usepackage[dvips]{epsfig}
\usepackage[dvips]{graphicx}
\usepackage{float}
\usepackage{units}
\usepackage{textcomp}

\usepackage{hyperref}
\usepackage{slashed}

\begin{document}


\title{Electronic properties of bilayer graphene catenoid bridge}


\author{J. E. G. Silva}
\affiliation{Universidade Federal do Cariri(UFCA), Av. Tenente Raimundo Rocha, \\ Cidade Universit\'{a}ria, Juazeiro do Norte, Cear\'{a}, CEP 63048-080, Brasil}

\author{J. Furtado}
\affiliation{Universidade Federal do Cariri(UFCA), Av. Tenente Raimundo Rocha, \\ Cidade Universit\'{a}ria, Juazeiro do Norte, Cear\'{a}, CEP 63048-080, Brasil}

\author{D. R. da Costa}
\affiliation{Universidade Federal do Cear\'a (UFC), Departamento de F\'isica,\\ Campus do Pici, Fortaleza - CE, C.P. 6030, 60455-760 - Brazil}

\author{T. M. Santhiago}
\affiliation{Universidade Federal do Cariri(UFCA), Av. Tenente Raimundo Rocha, \\ Cidade Universit\'{a}ria, Juazeiro do Norte, Cear\'{a}, CEP 63048-080, Brasil}

\author{Antonio C.A. Ramos}
\affiliation{Universidade Federal do Cariri(UFCA), Av. Tenente Raimundo Rocha, \\ Cidade Universit\'{a}ria, Juazeiro do Norte, Cear\'{a}, CEP 63048-080, Brasil}

\date{\today}

\begin{abstract}
We study the properties of an electron on a catenoid surface. The catenoid is understood as a realization of a bridge connecting two graphene layer by a smooth surface. The curvature induces a symmetrical reflectionless potential well around the bridge with one bound-state for $m=0$. For $m\neq 0$, a centrifugal potential barrier arises controlling the tunnelling between the layers. An external electric field breaks the parity symmetry and provides a barrier that controls the conductance from one layer to another. By applying a constant magnetic field the effective potential exhibits a confining double-well potential nearby the bridge. We obtain the corresponding bound states and study the effects of the curvature on the Landau levels. 
\end{abstract}


\maketitle

\section{Introduction}
\label{introduction}

In recent years two dimensional nanostructures, such as the graphene \cite{geim,novoselov,katsnelson}, nanotubes \cite{nanotube}  and the phosphorene \cite{phosphorene} has attracted attention due to their unusual properties. A two-dimensional single layer of carbon, known as graphene, exhibits no gap in the conductance band due to Dirac points yielding to a high conductance material \cite{katsnelson}. The bilayer graphene, on its turn, presents a quadratic dispersion relation which provides a gap in the conductance band. Such a gap allows the applications of the bilayer graphene in electronics \cite{katsnelson}.

Graphene properties can also be changed by the geometry of the layer. In conical layers the curvature at the tip induces topological phases \cite{furtado}.  Graphene strips in a helical present chiral properties \cite{dandoloff1,atanasovhelicoid,atanasov} known as chiraltronics, whereas M\"{o}bius-strip graphene is a topological insulator material \cite{mobius}. The effects of ripples \cite{contijo} and corrugated \cite{corrugated} surfaces upon electrons can also be described by geometric interactions. The curvature of the graphene sheet also produces effective interactions such as pseudomagnetic fields \cite{ribbons}.

The study of quantum mechanics on surfaces is a long-standing topic of debate. As pointed out by Dirac, the commutations relations between the position and momentum operators ought to be modified by the constrains which define the surface \cite{dirac}. By means of Feynman path integral, a classical particle minimally coupled to a surface using the induced metric induces a quantum potential proportional to the gaussian curvature \cite{dewitt,hjensen,cheng}. In another approach, defining the Laplacian operator in the tangent and normal coordinates and squeezing the particle on the surface, a geometric potential, known as the Da Costa potential is obtained \cite{costa}.  The geometric Da costa potential depends on the squared of the gaussian and the mean curvatures and yields to an attractive potential. This method also be extended to include external fields \cite{ferrari}, spin in a Pauli equation \cite{wang} and the Dirac equation on surfaces \cite{BJ}.

Curved graphene based structures can be used as new electronic devices. A bridge between two parallel graphene layers was proposed using a nanotube \cite{wormhole,picak}. A smooth bridge resembling a wormhole was proposed whose geometry induces an attractive potential nearby the throat brigde \cite{Dandoloff}. Another interesting bridge was proposed using a single catenoid surface \cite{dandoloff}. The catenoid is a surface of revolution of the catenary along some direction \cite{spivak}. It has the remarkable property of be a minimal surface, i.e., a surface of least area \cite{spivak}. As a result, the catenoid has negative gaussian curvature whereas the mean curvature vanishes at all points. The vanishing of the mean curvature ensures that the momentum normal to the surface vanishes identically \cite{wang2}. Minimal graphitic surfaces are hypothetical structures known to be stable \cite{terrones}. Besides the stability, the catenoid has the key property to be asymptotically flat, thereby describing the two layers far from the origin. Near the origin, the surface exhibits a smooth curved throat connecting the upper to the lower layer.

In this work we study the effects of the catenoid curvature, as well as background electric and magnetic fields, has upon a non-relativistic particle. In section \ref{section2}, by assuming the geometric Da Costa potential, we firstly adopt a coordinate system with enable us to define asymptotic free states, as expected from the flat geometry. Then, we analyse qualitatively the features of the effective potential, such as its behaviour with respect to the parity and time-reversal symmetry, due to the curvature, electric and magnetic fields. In section \ref{section3}, we employ numerical methods to obtain the energy spectrum and the eigenfunctions in order to study how the curvature and external fields modifies the spectrum and the bound states. Final remarks and perspectives are outlined in section \ref{remarks}.

\section{Electron on a catenoid surface}
\label{section2}

\begin{figure}
\label{catenoidsurface}
\includegraphics[scale=0.2]{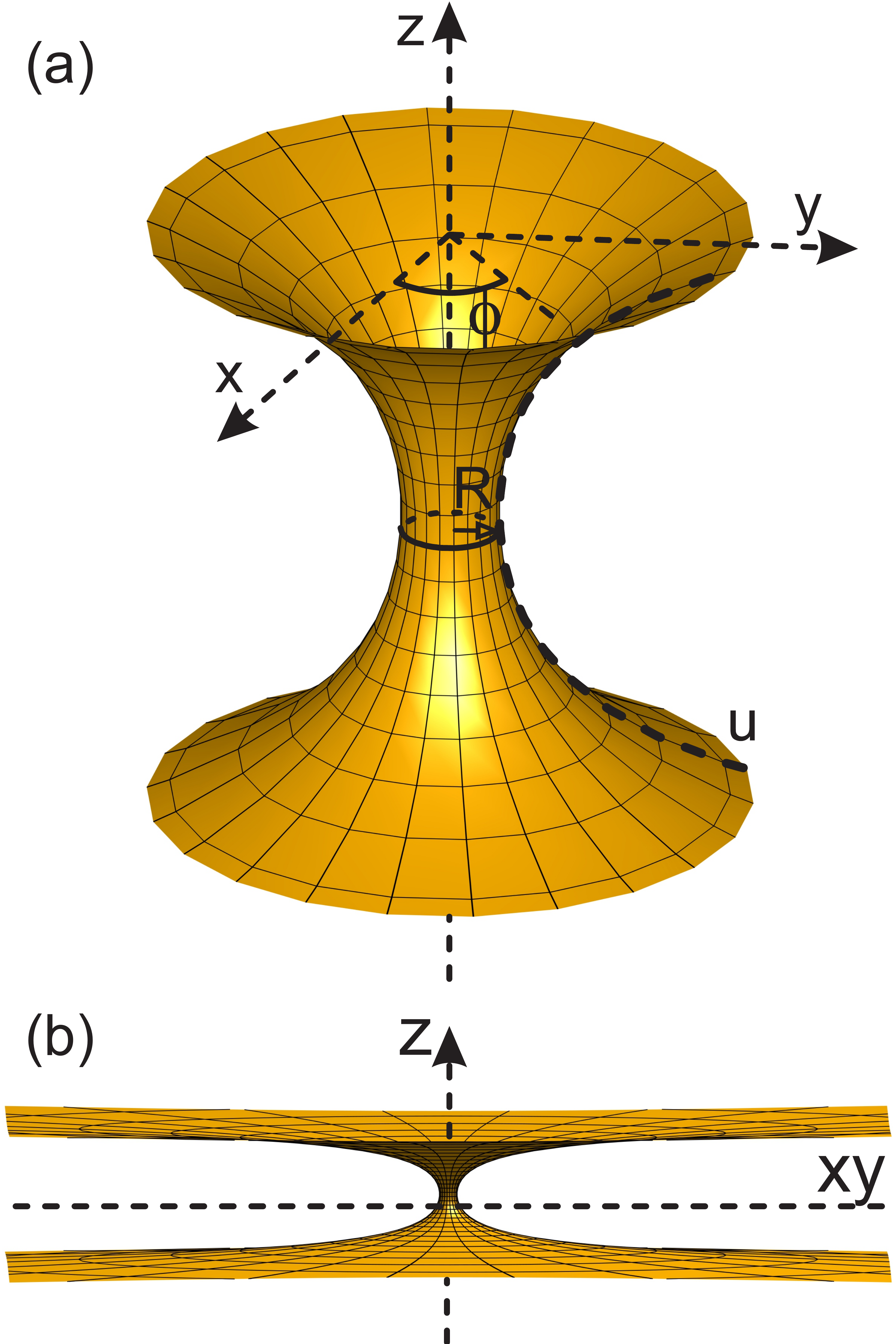}
\caption{Catenoid surface. Near the throat in fig. a) and seen at a large distance in fig. b).}
\end{figure}

In this section we introduce the geometry and the dynamics of the electron on the doublelayer catenoid bridge. As shown in fig.\ref{catenoidsurface}, the double layer graphene bridge is realized as a smooth minimal surface (least area) joining the two planes. Near the bridge throat, depicted in fig.\eqref{catenoidsurface}, the symmetry about the $z$ axis and the minimal radius $R$ are shown.

Consider an electron constrained to the catenoid surface and governed by the Hamiltonian \cite{ferrari,wang}
\begin{equation}
\label{hamiltonian}
\hat{H}=\frac{1}{2m^{*}}g^{ij}\hat{P}_i \hat{P}_j + V_{e} + V_{g},
\end{equation}
where $m^{*}$ is the electron effective mass, $\hat{P}_i :=-i \hbar \nabla_i - e A_k$ is the momentum operator of the electron minimally coupled to the magnetic field, $V_e$ is the electrostatic potential and $V_{g}$ is a confining potential which constrains the electron on the surface. The electron couples with the surface by means of the induced metric on the catenoid $g_{ij}$ and the covariant derivative $\nabla_i V^j:=\partial_i V^j + \Gamma^{j}_{ik}V^k $, where $ds^2=g_{ij}dx^{i}dx^{j}$ and $\Gamma^{j}_{ik}=\frac{g^{jm}}{2}(\partial_i g_{mk} + \partial_k g_{mi} - \partial_m g_{ik})$ is the Christoffel symbol \cite{spivak}. Throughout this work the index $i,j=\{1,2\}$ and stand label the catenoid coordinates. In addition, we consider a geometric potential, known as da Costa potential $V_{dC} = - \frac{\hbar^2}{2m^{*}}(H^2-K)$, where $H$ is the mean curvature and $K$ is the gaussian curvature \cite{costa}. 

In cylindrical coordinates the catenoid is parametrized as \cite{spivak}
\begin{equation}
\label{cilindricalcoordinates}
\vec{r}=R\cosh\left(z/R\right)\cos\phi\hat{i} +R\cosh\left(z/R\right)\sin\phi \hat{j} + z\hat{k},
\end{equation}
where $R$ the radius of the catenoid bridge, as shown in figure \ref{catenoidsurface}. We adopt a
coordinate system on the catenoid by taking the meridian $u=u(z)=R\sinh\left(z/R\right)$ and the parallel $\phi$. The meridian $u\in (-\infty, \infty)$ and the parallel $\phi\in [0,2\pi)$ coordinates cover the whole catenoid.
In this coordinate system, the interval reads $ds^{2}=du^2 + (R^2 +u^2)d\phi^2$,
and thence, the induced metric on the catenoid is $g_{uu}=1$ and $g_{\phi\phi}=R^2 +u^2$. The nonvanishing components of the Christoffel connection are $\Gamma_{\phi\phi}^{u}=-u$ and $\Gamma_{u\phi}^{\phi}=\frac{u}{R^2 + u^2}$.
Thus, the spinless stationary Schr\"{o}dinger equation is written as
\begin{eqnarray}
&-&\frac{\hbar^2}{2m^{*}}\Big[\Psi_{uu}+\frac{u}{R^2 + u^2}\Psi_{u}+\frac{1}{R^2 + u^2}\Psi_{\phi\phi}\Big]\nonumber\\
&+& ie\frac{\hbar}{2m^{*}}A^{j}\nabla_{j}\Psi + \frac{e^2}{2m^{*}}g_{jk}A^{j}A^k + \nonumber\\
&+& \left(V_e  - \frac{\hbar^2}{2m^{*}}\frac{R^2}{(R^2 + u^2)^2}\right)\Psi= \varepsilon\Psi,
\end{eqnarray}
where $\Psi_j=\frac{\partial \Psi}{\partial x^{j}}$. In the catenoid, the da Costa potential $V_{dC} =-\frac{\hbar^2}{2m^{*}}\frac{R^2}{(R^2 + u^2)^2}$ exhibits a parity-symmetrical potential well with respect to $u=0$. Asymptotically, the da Costa potential vanishes what reflects the asymptotic flat geometry of the catenoid. Furthermore, the curvature provides an attractive potential which tends to trap the electron on a ring around the origin.     
 
From the axial symmetry of the surface, the wave function has the periodic behaviour
\begin{equation}
\Psi(u,\phi)=\Phi(u)e^{im\phi},
\end{equation}
which yields the Schr\"{o}dinger equation along the meridian in the form
\begin{eqnarray}
\label{meridianschrodingerequation}
&-&\frac{\hbar^2}{2m^{*}}\Big[\Phi_{uu}+\frac{u}{R^2 + u^2}\Phi_{u}\Big] + ie\frac{\hbar}{2m^{*}}A^{j}\nabla_{j}\Phi + V_e\Phi\\
&+& \Bigg[\frac{e^2}{2m^{*}}A_{j}A^j + \frac{\hbar^2}{2m^{2}}\left(\frac{m^2}{R^2 + u^2}-\frac{R^2}{(R^2 + u^2)^2}\right)\Bigg]\Phi= \varepsilon\Phi\nonumber.
\end{eqnarray}
Note that the symmetry of the catenoid with respect to the $z$ axis induces a parity invariant centrifugal potential. Similar terms were found for the electron on a helicoid \cite{atanasovhelicoid,atanasov}.


\subsection{Geometry effects}
\label{geometry}
In order to understand the effects of the curved geometry of the catenoid upon the electron, let us study the properties of the wave function in absence of the electric and magnetic fields. 
The stationary Schr\"{o}dinger equation along the meridian \eqref{meridianschrodingerequation} reads
\begin{eqnarray}
\label{freenonhermitionequation}
\Phi_{uu}&+&\frac{u}{R^2 + u^2}\Phi_{u} + \nonumber\\
&-& \left(\frac{m^2}{R^2 + u^2} - \frac{R^2}{(R^2 + u^2)^2}\right)\Phi= -\frac{2 m^{*} \varepsilon}{\hbar^2}\Phi.
\end{eqnarray}
Note that the Eq.\eqref{freenonhermitionequation} exhibits parity and time-reversal invariance, as a result of the catenoid geometric symmetries, as we can see from the acting of parity $\mathcal{P}\hat{u}(z)\mathcal{P}=\hat{u}(-z)=-\hat{u}(z)$ and time reversal $\mathcal{T}\hat{u}(z)\mathcal{T}=\hat{u}(z)$ operators upon $u(z)$. Nonetheless, the first order derivative term renders the Hamiltonian non-Hermitian, since
\begin{eqnarray}
    \nonumber\left[\frac{i\hbar}{2m^{*}}\frac{\hat{u}}{R^2+\hat{u}^2}\hat{P}_u\right]^{\dagger}&=&-\frac{\hbar^2}{2m^*}\left[-\frac{2\hat{u}^2}{(R^2+\hat{u}^2)}+\frac{1}{R^2+\hat{u}^2}\right]\\
&&-\frac{i\hbar}{2m^{*}}\frac{\hat{u}}{R^2+\hat{u}^2}\hat{P}_u,
\end{eqnarray}
where $\hat{P}_u := -i\hbar \partial_u$ and $\hat{u}$ are indeed Dirac hermitean. The non-hermiticity of the free electron Hamiltonian is not a problem, since the space-time reflection symmetry is preserved, the spectrum of the eigenvalues of the Hamiltonian is completely real \cite{Bender, Bender2}. Besides, there is an Hermitean equivalent Hamiltonian that can be achieved by a simple changing of variables. Considering the change on the wave function 
\begin{equation}
    \Phi(u)=\frac{1}{(R^2+u^2)^{1/4}}y(u),
\end{equation}
leads to an one dimensional Hermitian Schr\"{o}dinger equation
\begin{equation}\label{modifiedschrodingerequation}
    -y_{uu}+V_{eff}(u)y=\frac{2m^{*}\varepsilon}{\hbar^2}y,
\end{equation}
whose effective potential is given by
\begin{equation}
\label{freeeffectivepotential}
V_{eff}=\frac{\hbar^2}{2m^{*}}\left[\frac{(2m^2+1)}{2(R^2+u^2)}-\frac{(3u^2+4R^2)}{4(R^2+u^2)^2}\right].
\end{equation}
Therefore, the dynamics of an electron on a catenoid with Hamiltonian \eqref{hamiltonian} is Hermitian equivalent to 
an electron under the action of the effective potential in Eq.\eqref{freeeffectivepotential}. 
The equivalence between a non-hermitean Hamiltonian which preserves the $\mathcal{P}$ $\mathcal{T}$ symmetry and an Hermitean one was already discussed by a number of authors \cite{Jones, Andrianov1, Andrianov2}. 

The Schr\"{o}dinger equation $\eqref{modifiedschrodingerequation}$ allow us to obtain the asymptotic free states. Indeed, for 
$u\rightarrow \infty$, the potential $\eqref{freeeffectivepotential}$ vanishes and the respective solution are
\begin{equation}
y(u)=A\cos(k u + \varphi)
\end{equation}
where $k^2=\frac{2 m^{*} \varepsilon}{\hbar^2}$. The same result can be obtained from Eq.\eqref{freenonhermitionequation} and it reflects the asymptotic flatness of the catenoid.

\begin{figure}[h!]
\begin{center}
\includegraphics[scale=0.5]{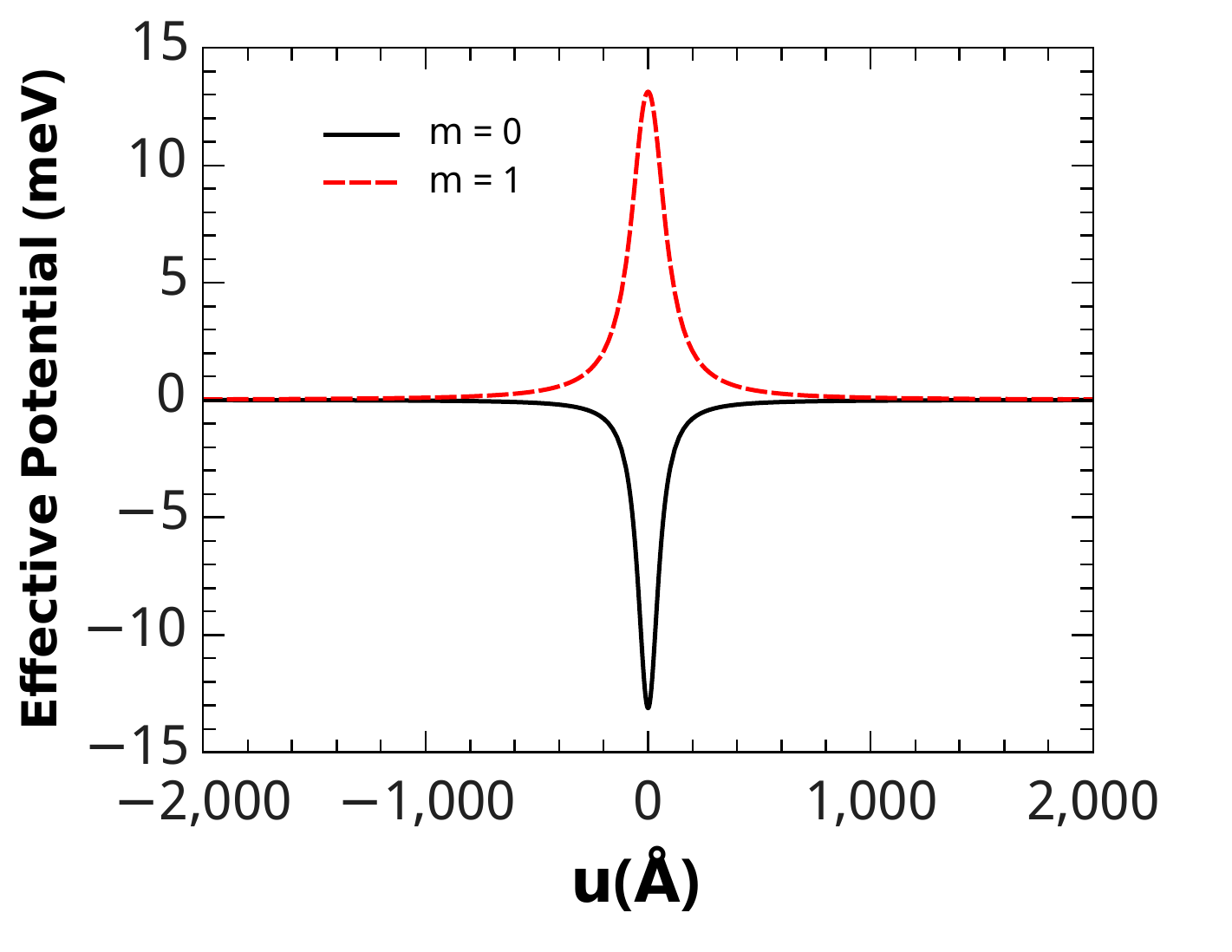}  
\caption{The effective potential for $R=70\AA$ and $E=B=0$. The thin black line correspond to $m=0$ whereas the dashed red line represents $m=\pm 1$.}
\label{FIG1}
\end{center}
\end{figure}

The effects of the curvature on the electron can be seen by the features of the effective potential. 
In the fig.\ref{FIG1} we present the effective potential for $R=70\AA$ and we adopt the effective electron mass in the graphene, $m^{*}=0.03m_{0}$ \cite{grafe}. 

For $m=0$ the centrifugal potential is absent and the curvature produces the potential well around the origin of the catenoid (solid black line). This symmetric potential has a reflectionless shape and then, all asymptotic free state $k^2>0$ approaching the catenoid bridge near the origin will undergo complete transmission \cite{lekner}. The curvature increases the depth of the potential which tends to delta-type potential as $R\rightarrow 0$.

For $m=\pm 1$, the centrifugal term dominates over the attractive term and the effective potential produces a symmetric barrier around the origin (dashed red line). Therefore, an asymptotic free electron can be partially transmitted and reflected by the bridge curvature.

In the limit as $R\rightarrow 0$, the stationary wave function has the form $\Psi_{k,m}(u) = N J_{m}(ku)e^{l\theta}$, where $J_m$ is the Bessel function of first-kind. A similar wave function was found by considering a Dirac fermion in a graphene bridge build using two parallel layers and one nanotube as the bridge \cite{picak}.  Thus, even for a infinitely thin bridge, an asymptotic free $m=0$ state can tunnel from the upper (lower) layer to the lower (upper) layer through the bridge. For $m\neq 0$ the wave function vanishes at the origin, as expected from the potential barrier.

\subsection{Constant electric field}
\label{electric}

In this section we consider the electron under the action of an external electric field pointing in the positive $z$ direction, i.e., $\vec{E}=E\hat{k}$. Projection the electric field on the catenoid, we obtain $\vec{E}_u =E\frac{R}{\sqrt{R^2 + u^2}} \hat{e}_u$, where $\hat{e}_u = \frac{\partial \vec{r}}{\partial u} = \frac{1}{\sqrt{R^2 + u^2}}(u \hat{e}_\rho + R \hat{k})$ and $E_\phi = \vec{E}\cdot \hat{e}_\phi =0$. The electrostatic potential energy upon the electron on the catenoid has the form
\begin{equation}
\label{elecpotentialenergy}
V_e (u) =e E R \sinh^{-1}(u/R). 
\end{equation} 
Therefore, the effective potential becomes 
\begin{eqnarray}
\label{6a}
V_{eff}&=&\frac{\hbar^2}{2m^{*}}\left[\frac{(2m^2+1)}{2(R^2+u^2)}-\frac{(3u^2+4R^2)}{4(R^2+u^2)^2}\right]\nonumber\\
&+& e E R \sinh^{-1}(u/R). 
\end{eqnarray}

\begin{figure}[h!]
\begin{center}
\includegraphics[scale=0.5]{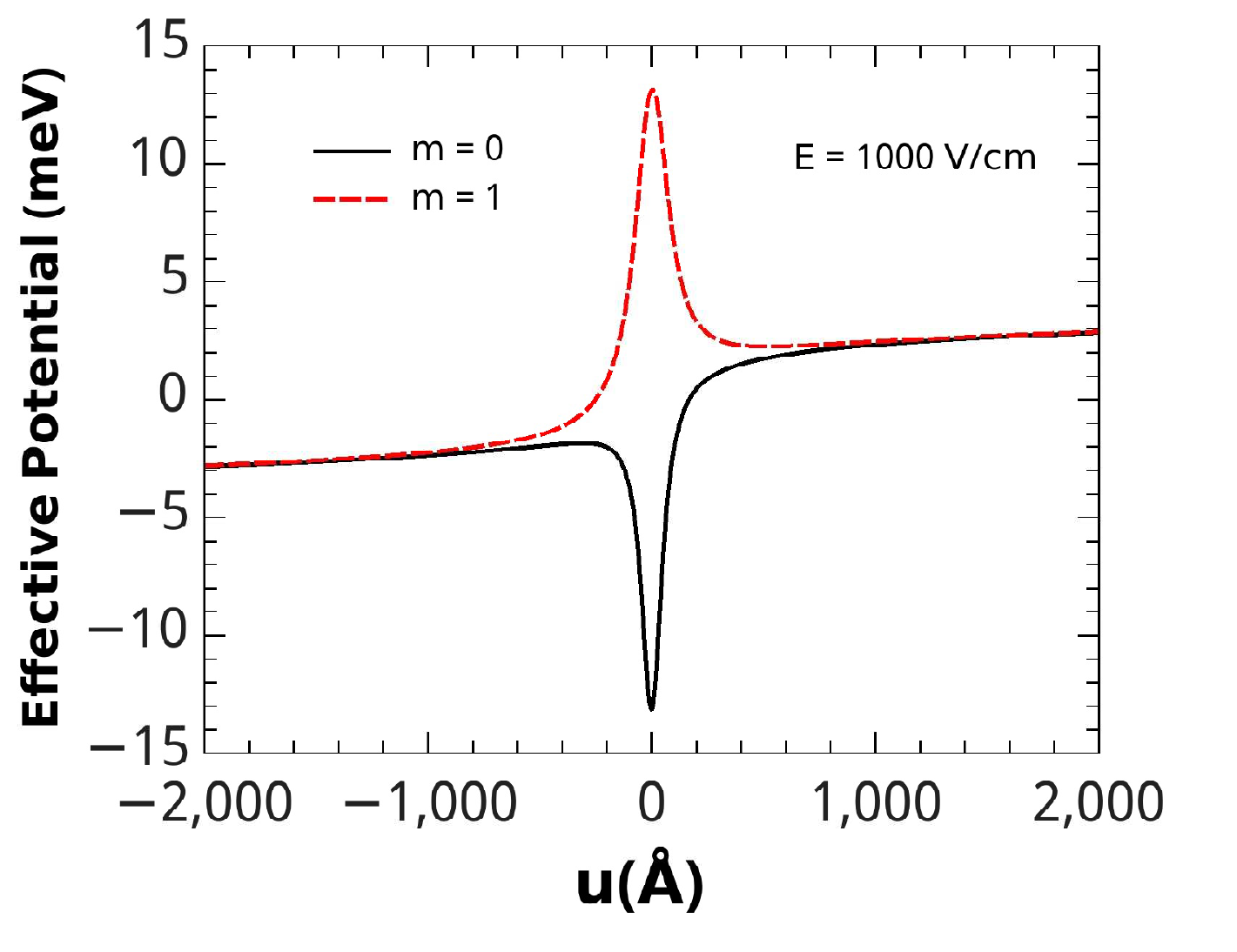} 
\caption{The effective potential for $R=70$\AA, $E=1$kV/cm and $B=0$. The thin black line correspond to $m=0$ whereas the dashed red line represents $m=\pm 1$.}
\label{FIG2}
\end{center}
\end{figure}

The background electric field breaks the parity symmetry with respect to the coordinate $u$, since $\mathcal{P}V_e(\hat{u})\mathcal{P}=-V_e(\hat{u})$. This effect is shown in the figs.\ref{FIG2} in which an electric field, of $1$kV/cm, is applied over the catenoid. The asymmetry on the potential produce a diode-like effect increasing the energy required to tunnel through the bridge. However the presence of an external electric field has no effect under time reversion, thereby the hamiltonian remains $\mathcal{T}$-symmetric.
 
 
\subsection{Constant magnetic field}
\label{magnetic}

Let us consider the properties of the electron subjected to an external magnetic field $\vec{B}=B\hat{k}$. The vector potential $\vec{A}$ has the form $\vec{A}=\frac{1}{2}\vec{B}\times \vec{r}=\frac{B}{2}\sqrt{R^2 + u^2}\hat{e}_\phi$, where $A^\phi=A^\phi (u)=\frac{B}{2}\sqrt{R^2 + u^2}$ and $A^u =0$. The effective potential becomes 
\begin{eqnarray}
\label{7a}
V_{eff}&=&\frac{\hbar^2}{2m^{*}}\Bigg[\frac{(2m^2+1)}{2(R^2+u^2)}-\frac{(3u^2+4R^2)}{4(R^2+u^2)^2}-\frac{eBm}{\hbar}\nonumber\\
&+& \frac{e^2 B^2}{4\hbar^2}(R^{2}+u^{2})\Bigg] +eE R \sinh^{-1}(u/R).
\end{eqnarray}

\begin{figure}[ht!]
\begin{center}
\includegraphics[scale=0.5]{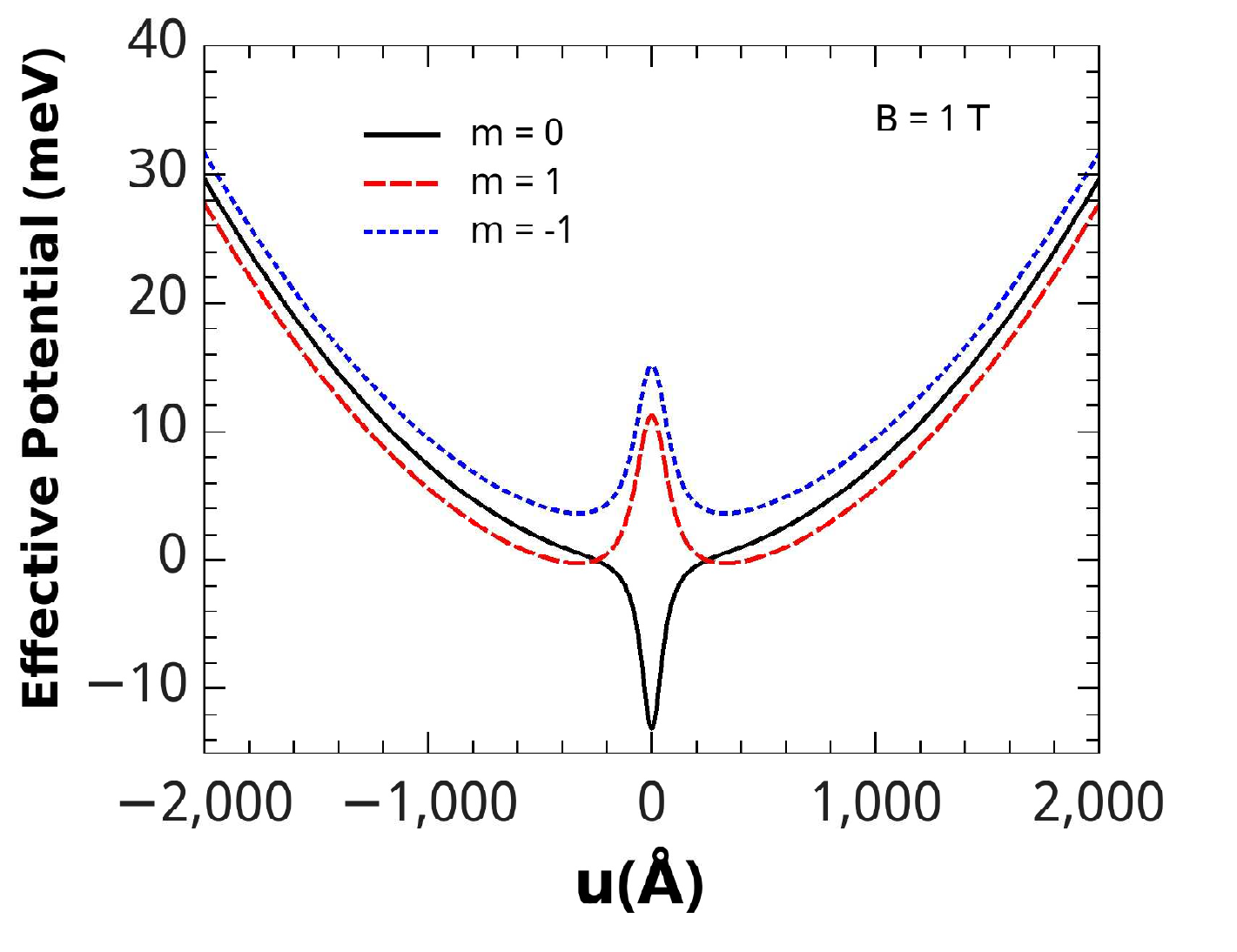}   
\caption{The effective potential for $R=70$\AA, $E=0$ and $B=1$ T.. The thin black line correspond to $m=0$ whereas the dashed red line represents $m=1$ and the blue dotted line stands for $m=-1$. }
\label{FIG3}
\end{center}
\end{figure}
The profile of the effective potential in the absence of electric field  $E=0$ kV/cm and using $B=1$ T, $R=70\AA$ is shown in the figs.\ref{FIG3}. We increased the domain of the figures, to $-2000$\AA$<u<2000$\AA, in order to improve their visualization. For ($m=0$) the potential exhibits a deep well around the origin and a parabolic potential for larger $u$. For $m\neq 0$, the magnetic field produces a symmetric double well potential with a barrier around the origin. Moreover, the term linear in $m$ in the effective potential Eq.\eqref{7a} produces chiral effects by breaking the symmetry $m \rightarrow -m$ as well as the symmetry under temporal reversion. The parity symmetry remains unchanged in the presence of a constant magnetic field.

\begin{figure}[ht!]
\begin{center}
\includegraphics[scale=0.5]{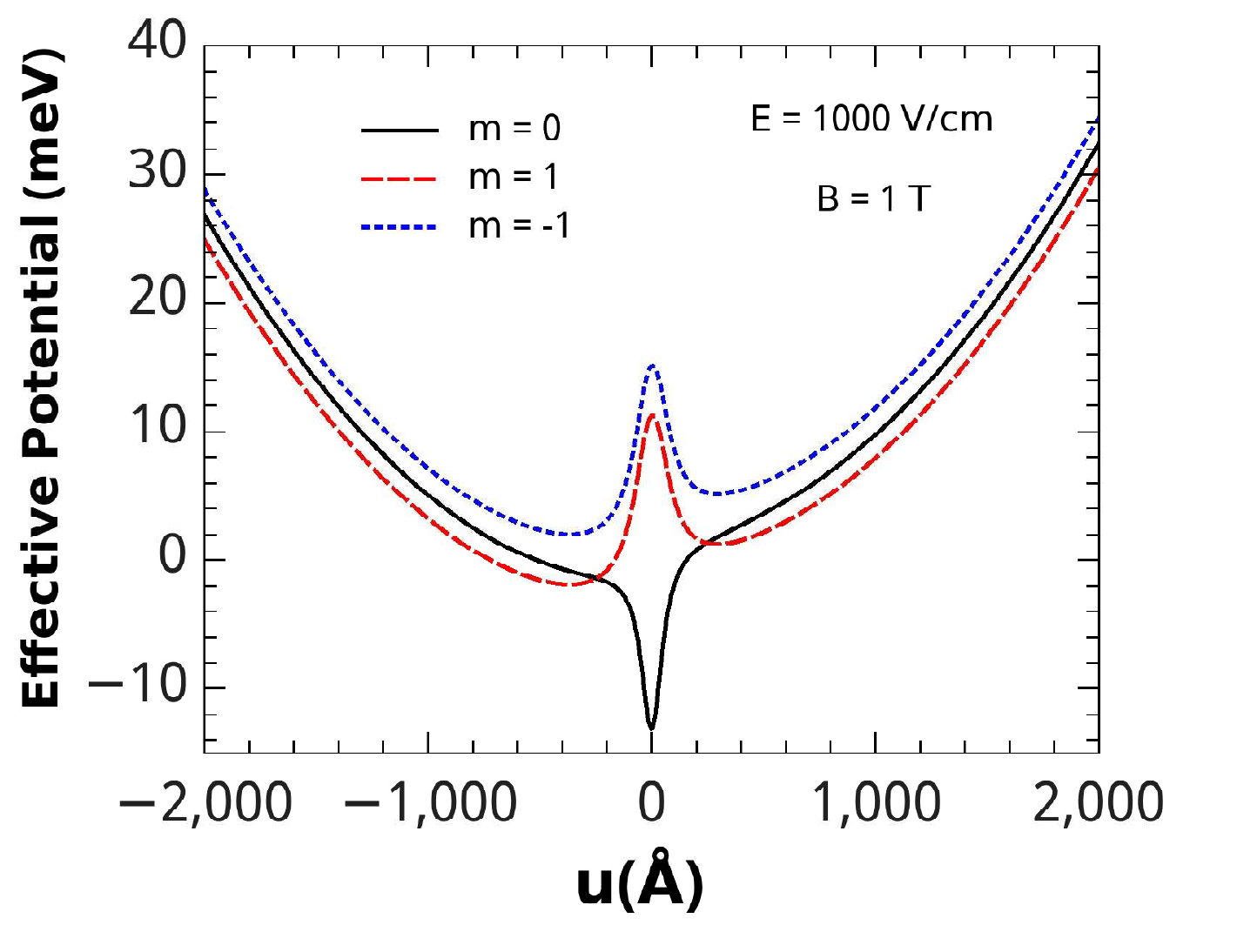}    
\caption{The effective potential for $R=70$\AA, $E=1$ kV/cm and $B=1$ T.. The thin black line correspond to $m=0$ whereas the dashed red line represents $m=1$ and the blue dotted line stands for $m=-1$}
\label{FIG4}
\end{center}
\end{figure}

By applying an electric field $E=1$ kV/cm the potential exhibits an asymmetric double potential shown in figs.\ref{FIG4}.
This asymmetry can be used to trap the electrons in ring-like regions near the bridge on the upper or lower layers.


\section{Bound states}
\label{section3}

In the previous section we studied qualitatively the features of the electron on the catenoid by examining the characteristics of the potential. In this section we obtain the bound states nearby the bridge due to the geometry, electric and magnetic fields.

Firstly, let us rewritte the Schr\"{o}dinger equation \eqref{meridianschrodingerequation} as \cite{Ramos11}     
\begin{equation}
-\frac{\hbar^2}{2m^{*}}\frac{1}{\sqrt{(R^2 + u^2)}}\frac{d}{du}(\sqrt{R^2 + u^2}\Phi_u ) + V_{eff}\Phi = \varepsilon\Phi.
\label{3a}
\end{equation}
We solve numerically the eq.\eqref{3a} using the $R=70$ \AA, and $m^{*}=0.03 m_{0}$, that is, the effective mass of the carbon.  

\begin{figure*}[ht!]
\begin{center}
\includegraphics[scale=0.5]{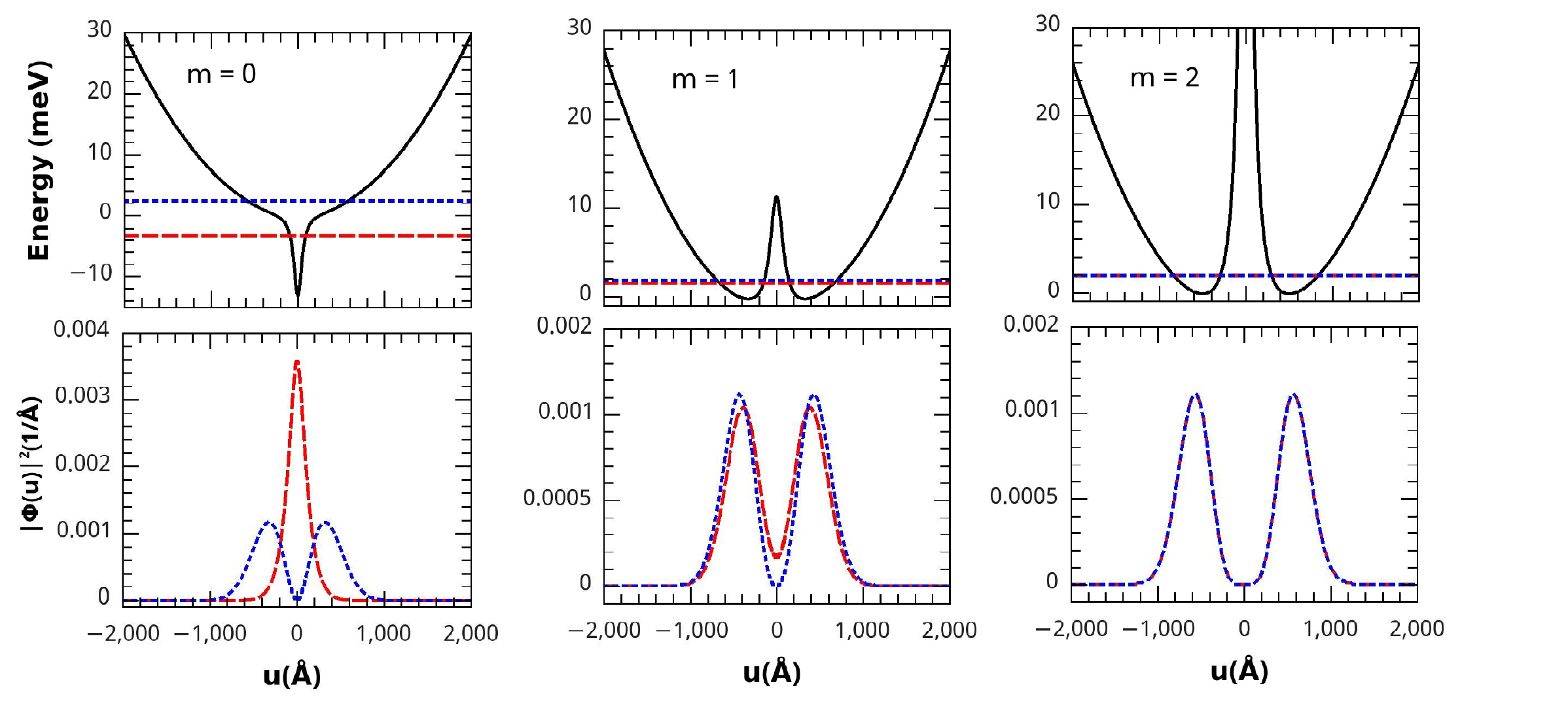}   
\caption{This figure shows the energy levels plotted together with the effective potential (solid black line) and the probability density function for $m=0,1,2$, $R=70\AA$, $E=0$V/cm and $B=1$T. The red dashed lines and the dotted blue lines correspond to the first and second energy levels.}
\label{FIG6}
\end{center}
\end{figure*}

In fig.\eqref{FIG6} we show the energy spectrum (Landau levels) for the $m=0$, $m=1$ and $m=2$ with their respective probability distribution in the absence of the electric field. For $m=0$ the ground state (red dashed line) has a gaussian profile localized around the origin (cathenoid neck) whereas the first excited state (blue dotted line) vanishes at the origin and has two symmetric peaks around the origin. The ground state bears a resemblance with the bound state due to the only the geometric potential. For $m=1,2$, the centrifugal term shifts the bound states from the origin and produces bound ring-like states at the upper and lower layers around the cathenoid neck.

\begin{figure*}[ht!]
\begin{center}
\includegraphics[scale=0.5]{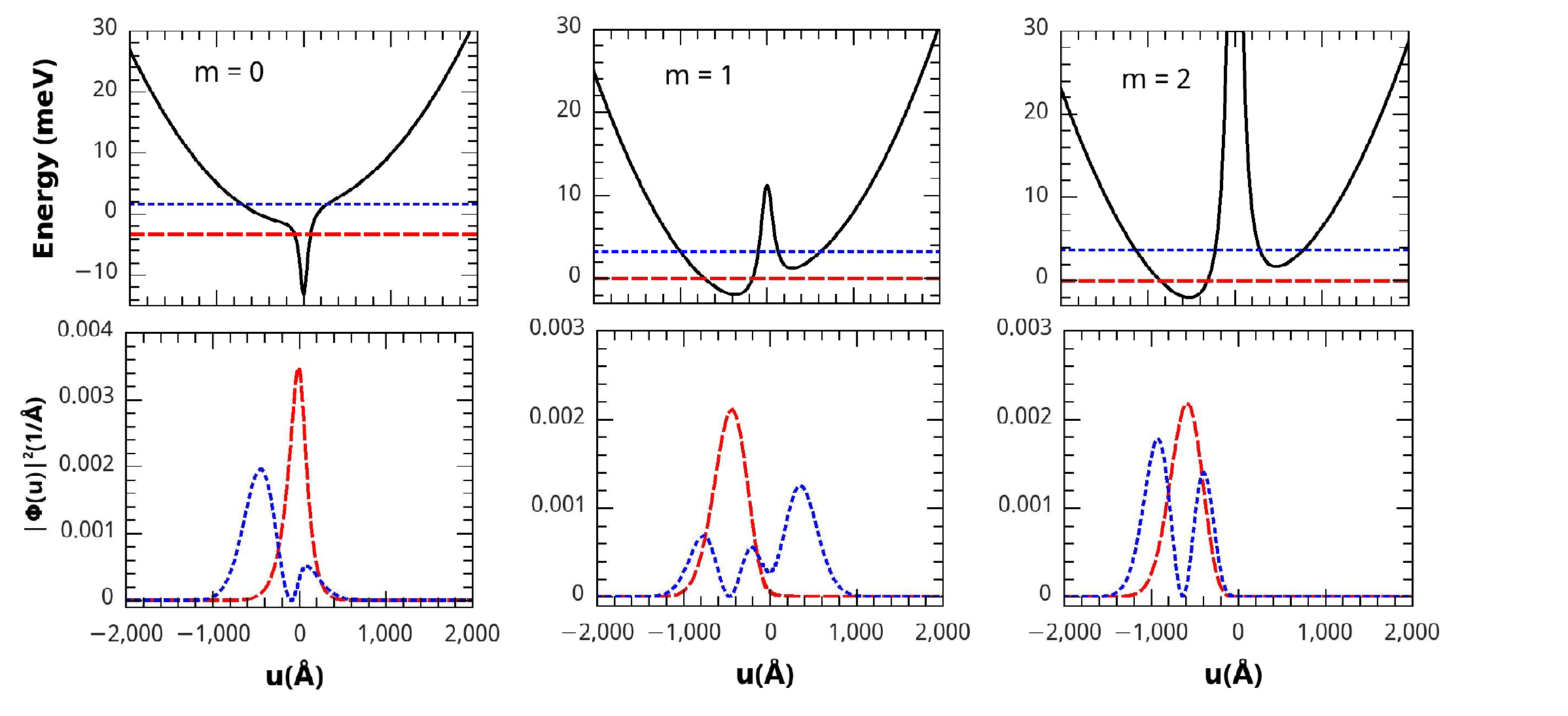}   
\caption{This figure shows the energy levels plotted together with the effective potential (solid black line) and the probability density function for $m=0,1,2$, $R=70\AA$, $E=1$kV/cm and $B=1$T. The red dashed lines and the dotted blue lines correspond to the first and second energy levels.}
\label{FIG7}
\end{center}
\end{figure*}

An external electric field $E=1$kV/cm the produces an asymmetric effective potential which tends to trap the electron on the lower layer, as shown in Fig.\eqref{FIG7}. For $m=0$ the asymmetry effect is stronger upon the first excited state. The localization of the electron on the lower layer increases with $m$. For $m=2$ the first excited state is localized in two ring-like regions.

\section{Final Remarks and perspectives}
\label{remarks}

We explored the electronic properties of double-layer graphene bridge realized as a catenoid surface. A Hamiltonian quadratic in the momentum was considered, whose square is measured using the induce metric on the surface. That prescription ensures the independence of the kinetic term with respect to the coordinate system. Along the meridian direction, we obtain a one dimensional Hamiltonian with a non-Hermitian term steaming from the momentum correction due to the Christoffel symbols. By a wave function redefinition, we obtained a Hermitian Hamiltonian whose effective potential carries all the curvature effects. In addition to this minimal coupling, we considered a potential depending on the curvature, known as the da Costa potential \cite{costa}.

The geometry induces an attractive potential around the origin whereas the axial symmetry provides a centrifugal barrier. Both potentials vanish asymptotically reflecting the asymptotic flat geometry of the catenoid. Accordingly, free states are easily defined for $u\rightarrow\infty$. As the free $m=0$ states approach the bridge they "interact" with the surface with a reflectionless potential allowing complete transmission. Nevertheless, the barrier "felt" by the $m\neq 0$ states produces transmitted and reflected states. A detailed analysis of the scattered states by the centrifugal barrier around the bridge is an important development of the present work.

Moreover, we studied the effects of external field upon the electron on the surface. It turns out that an electric field in the $z$ direction breaks the parity symmetry between the upper and the lower layers, producing a diode-like step potential \cite{zhu}. The effects of this asymmetry on the conducting electrons is another worthy perspective. An external magnetic field yields a parabolic well for $m=0$ and a  double well potential for $m\neq 0$.

We carried out numerical analysis to find bound states trapped around the bridge due to the curvature and the external fields. The $m=0$ geometric potential well allows one bound state, a characteristic feature of the reflectionless potentials \cite{lekner}. The curvature also modifies the Landau levels produced by an external magnetic field. The parabolic potential for $m=0$ is modified due to the curved geometry, thereby altering the separation between the energy eigenvalues. The double-well potential for $m\neq 0$ enables bound states on each minimum. These bound states represents electrons trapped in ring-like regions above or below the bridge. An interesting effect to be investigated is the possible tunnelling of electron between these two minimum.  By applying an external electric field, it is possible to produce a difference between the upper and lower minima. Further, the shift in the Landau levels due to the curvature seems a promising feature for quantum dots applications. 

Besides the scattering analysis, the effects of the electron spin described by the Pauli equation is an import improvement for future investigations. The thermodynamic properties of an electron gas on this surface is another noteworthy perspective.



\begin{thebibliography}{99}

\bibitem{geim} A. K. Geim, K. S. Novoselov, Nature Materials 6, 183 (2007).

\bibitem{novoselov} A. H. Castro Neto, F. Guinea, N. M. R. Peres, K. S. Novoselov, A. K. Geim, Rev. Mod. Phys. 81, 109 (2009).

\bibitem{nanotube} S. Berber, Y.K. Kwon and D. Tomanek, Phys. Rev. Lett. {\bf 84}, 4613 (2000).

\bibitem{phosphorene} A. Carvalho, M. Wang, X. Zhu, A. S. Rodin, H. Su, A. H. Castro Neto, Nat. Rev. Mat. {\bf 1}, 11 (2016).  

\bibitem{katsnelson} M. Katsnelson, Graphene: Carbon in two dimensions, Cambridge University Press, Cambridge, (2012).

\bibitem{furtado} C. Furtado , F. Moraes , A.M. de M. Carvalho, Phys. Lett. A {\bf 372}, 5368, (2008).

\bibitem{dandoloff1} R. Dandoloff, T.~T.~Truong, Phys.~Lett.~A {\bf 325}, 233 (2004).


\bibitem{atanasovhelicoid} V. Atasanov, R. Dandoloff and A. Saxena Phys. Rev. B 79, 033404 (2009).
 

\bibitem{atanasov} V. Atasanov, A. Saxena, Phys. Rev. B 92, 035440 (2015).


\bibitem{mobius} Z. L. Guo, Z. R. Gong, H. Dong, C. P. Sun, Phys.~Rev.~B {\bf 80}, 195310 (2009).


\bibitem{contijo} F. de Juan, A. Cortijo, M. A. H. Vozmediano, Phys. Rev. B 76, 165409, (2007).


\bibitem{corrugated} V. Atasanov, A. Saxena, Phys. Rev. B {\bf 81}, 205409 (2010).


\bibitem{ribbons} F. Guinea, A. K. Geim, M. I. Katsnelson, and K. S. Novoselov, Phys. Rev. B 81, 035408 (2010).

\bibitem{dirac} P. A. M. Dirac, The principles of quantum mechanics, (Oxford University Press, New York, 1947).

\bibitem{dewitt} B. S. DeWitt, Rev. Mod. Phys. {\bf 29}, 377 (1957).

\bibitem{hjensen} H. Jensen, H. Koppe, Ann. Phys. {\bf 63}, 586 (1971).

\bibitem{cheng} K.~S.~Cheng, Jorn.~Math.~Phys.~{\bf 13}, 1793 (1972).




\bibitem{costa} R. C. T. da Costa, Phys. Rev. A 23, 4, (1981).

\bibitem{ferrari} G. Ferrari, G. Cuoghi, Phys. Rev. Lett. 100, 230403 (2008).

\bibitem{wang}  Y. Wang, L. Du, C. Xu, X. Liu and H. Zong Phys. Rev. A90, 042117 (2014).


\bibitem{BJ} M. Burgess, B. Jensen, Phys. Rev. A 48, 3, (1993).

\bibitem{wormhole} J.~Gonz\'{a}lez, J.~Herrero, Nucl.~Phys.~B {\bf 825}, 426 (2010).

\bibitem{picak} R.~Pincak, J.~Smotlacha, Quantum Matter {\bf 5}, 114 (2016). 

\bibitem{Dandoloff} R. Dandoloff, Phys. Lett. A, {\bf 373} (2009).

\bibitem{dandoloff} R. Dandoloff, A. Saxena, B. Jensen, Phys. Rev. A 81, 014102 (2010).

\bibitem{spivak} M. Spivak, A comprehensive introduction to differential geometry, Publish or Perish, Houston, (1999).

\bibitem{wang2} Y. Wang, H. Jiang, H. Zong, Phys.~Rev.~A {\bf 96}, 022116 (2017).


\bibitem{terrones} H. Terrones, A. L. Mackay, Chem. Phys. Let. 207, 45, (1993).

\bibitem{tagami} M. Tagami, Y. Liang, H. Naito, Y. Kawazoe, M. Kotani, Carbon {\bf 76}, 266 (2004).




\bibitem{grafe} J. Li, L. Z.Tan, K. Zou, A. A. Stabile, D. J. Seiwell, K. Watanabe, T. Taniguchi, Steven G. Louie, and J. Zhu, Phys. Rev. B 94, 161406R (2016).

\bibitem{Ramos11} A. C. A. Ramos, G. A. Farias and N. S. Almeida, Physica E 43, 1878 (2011).















\bibitem{lekner} J. Lekner, Am.~J.~Phys. {\bf 75}, 1151 (2007).



\bibitem{Bender} C.~M.~Bender, Rept.\ Prog.\ Phys.\  {\bf 70}, 947 (2007)

\bibitem{Bender2} C.~M.~Bender and S.~Boettcher, Phys.\ Rev.\ Lett.\  {\bf 80}, 5243 (1998)
 
\bibitem{Jones} H.~F.~Jones and J.~Mateo,  Phys.\ Rev.\ D {\bf 73}, 085002 (2006)

\bibitem{Andrianov1} A.~A.~Andrianov, Phys.\ Rev.\ D {\bf 76}, 025003 (2007)

\bibitem{Andrianov2} A.~A.~Andrianov, Annals Phys.\  {\bf 140}, 82 (1982).
  
\bibitem{zhu} Z. Zhu, S. Joshi, S. Grover, G. Moddel, J. Phys. D: Appl. Phys. {\bf 46} (2013).


\end{thebibliography}
\end{document}